\documentclass[aps,superscriptaddress,a4paper,nofootinbib,floatfix,twocolumn,10pt,pra]{revtex4-1}
\usepackage{graphicx,color,amsmath,amssymb,enumerate,verbatim,graphicx,color,hyphenat, tikz, standalone, mathtools, url}\usetikzlibrary{shapes, arrows}
\usepackage[inline]{enumitem}
\usepackage[normalem]{ulem}

\usepackage[ocgcolorlinks,colorlinks=true,linkcolor=blue,citecolor=red]{hyperref}
\usepackage{multirow}

\newtheorem{theo}{Theorem}
\newtheorem{defin}{Definition}

\newcommand{\tr}{\mathrm{tr}}
\newcommand{\ra}{\rangle}
\newcommand{\la}{\langle}
\newcommand{\bra}[1]{\langle #1|}
\newcommand{\ket}[1]{|#1\rangle}

\newcommand{\braket}[3]{\langle #1 | #2 | #3 \rangle}

\newcommand{\batt}{\mathrm{batt}}
\newcommand{\sys}{\mathrm{s}}
\newcommand{\total}{\mathrm{total}}
\newcommand{\s}{\mathrm{s}}
\renewcommand{\b}{\mathrm{b}}
\renewcommand{\sb}{\mathrm{sb}}

\newcommand{\Mod}[1]{\left|#1\right|}

\def\phi{\varphi}
\DeclarePairedDelimiter\ceil{\lceil}{\rceil}
\DeclarePairedDelimiter\floor{\lfloor}{\rfloor}

\begin{document}
\title{Thermodynamics of quantum systems with multiple conserved quantities}
\author{Yelena Guryanova}\affiliation{H. H. Wills Physics Laboratory, University of Bristol$\text{,}$ Tyndall Avenue, Bristol, BS8 1TL, United Kingdom}
\author{Sandu Popescu} \affiliation{H. H. Wills Physics Laboratory, University of Bristol$\text{,}$ Tyndall Avenue, Bristol, BS8 1TL, United Kingdom}
\author{Anthony J.~Short}\affiliation{H. H. Wills Physics Laboratory, University of Bristol$\text{,}$ Tyndall Avenue, Bristol, BS8 1TL, United Kingdom}
\author{Ralph Silva}\affiliation{D\'epartement de Physique Th\'eorique, Universit\'e de Gen\`eve, 1211 Gen\`eve, Switzerland}
\affiliation{H. H. Wills Physics Laboratory, University of Bristol$\text{,}$ Tyndall Avenue, Bristol, BS8 1TL, United Kingdom}
\author{Paul Skrzypczyk}\affiliation{H. H. Wills Physics Laboratory, University of Bristol$\text{,}$ Tyndall Avenue, Bristol, BS8 1TL, United Kingdom}

\begin{abstract}

We consider a generalisation of thermodynamics that deals with multiple conserved quantities at the level of individual quantum systems. Each conserved quantity, which, importantly, need not commute with the rest, can be extracted and stored in its own battery. Unlike in standard thermodynamics, where the second law places a constraint on how much of the conserved quantity (energy) that can be extracted, here, on the contrary, there is no limit on how much of any individual conserved quantity that can be extracted. However, other conserved quantities must be supplied, and the second law constrains the combination of extractable quantities and the trade-offs between them which are allowed. We present explicit protocols which allow us to perform arbitrarily good trade-offs and extract arbitrarily good combinations of conserved quantities from individual quantum systems.

\end{abstract}

\maketitle

\section{Introduction}

Thermodynamics is one of the most successful theories of nature that we have. Since its inception almost 200 years ago it has survived the transition from classical mechanics to relativistic and quantum mechanics with its conceptual basis unchanged. The realm of thermodynamics has also been considerably extended, with recent years witnessing the extension of thermodynamics from dealing with macroscopic systems to individual quantum systems and black holes. From its earliest days, thermodynamics was also generalised to deal not only with energy, but various conserved quantities, introducing grand canonical ensembles, chemical potentials, etc. Again, its conceptual basis remained unchanged. 

Here we present a generalised version of thermodynamics which deals with individual quantum systems and multiple -- commuting or non-commuting -- conserved quantities. What we will show, is that unlike standard thermodynamics, where the second law constrains how much of the conserved quantity (energy) can be extracted from a non-equilibrium system, in the form of work, here there is no constraint on the amount of a single conserved quantity that can be extracted. In fact we can extract as much of any individual conserved quantity as we like, if we supply an appropriate amount of other conserved quantities. What the second law constrains is the \emph{combination} of conserved quantities that can be extracted -- that is, the second law is seen to limit the trade-off of extractable quantities, which for the standard case of a single conserved quantity reduces to constraining how much can be extracted. At the same time, this generalised version of thermodynamics suggests that it may be worthwhile revisiting the basic concepts of the subject. Indeed, to understand the above phenomena we present an alternative viewpoint, which reinterprets some of the standard thermodynamics quantities, and is perhaps more natural and compelling.

There are many precursors to the results presented here. The idea of the grand canonical ensemble, where not only energy, but also the number of particles, is a conserved quantity, goes all the way back to Gibbs \cite{Gibbs1902}. A milestone was the work in 1957 of Jaynes \cite{jaynes1,Jaynes1957}, who, coming from a Bayesian perspective, suggested the generalisation of thermodynamics to arbitrary conserved quantities through the principle of maximum entropy. The idea of the `generalised Gibbs ensemble' is by now commonly used in quantum statistical mechanics, see e.g. \cite{Kollar2011,Caux2012,Dora2012,Pozsgay2013}. More recently, \cite{joan2,joan1} considered Landauer erasure given access to a `angular momentum bath' instead of a thermal bath, and demonstrated that information can be erased without an energy cost, provided the analogous cost is paid in angular momentum.

Very recently, there has been much renewed interest in the foundations of thermodynamics coming from the field of quantum information. On the one hand, the so called `single shot information theory', which was developed initially to study finite size effects in quantum cryptography, has proven useful for studying finite size effects and fluctuations in quantum thermodynamics, which has lead to the development of `single shot quantum thermodynamics' \cite{DahRenRie11,RioAbeRen11,Egloff2015,Faist2015,Halpern2015,Woods2015,Weilenmann2015}. On the other hand, inspired by so called `resource theories' which have proved to be very powerful for studying quantum information tasks, such as the theory of entanglement \cite{BenBerPop96}, purity \cite{HorHorOpp03} or asymmetry \cite{Gour2008}, the `resource theory of quantum thermodynamics' \cite{Brandao2013a} was developed which, in combination with single shot framework, has generated a lot of interest and already shown itself to be a fruitful approach to quantum thermodynamics \cite{JanWocZei00,HorOpp11,Aberg2013a,paul,secondlaws,Ng2015,Renes2014,Faist2015a,Lostaglio2015,Lostaglio2015a,Cwiklinski2015,Lostaglio2015b,Korzekwa2015,Alhambra2015,Perry2015} Our work proceeds along the lines of single-shot thermodynamics and resource theories, where a number of initial results concerning the thermodynamics of multiple conserved quantities were presented in \cite{nicole1,nicole2}.

The paper is organised as follows: We first summarise the main results in Sec.~II. We then review the notion of a generalised thermal state and introduce the free entropy in Sec.~III, and follow on by discussing the conceptual aspects of this work in Sec.~IV. In Sec. V we introduce more formally the framework. In Sec. VI we state and prove the second law. Then in Sec. VII and VIII we consider the case of commuting, and non-commuting observables respectively. Finally in Sec. IX we conclude.

\section{Main results}
Here we consider the standard general framework of thermodynamics, that consists of a thermal bath, an external system out of equilibrium with respect to the bath, and a number of batteries, in which we will store various conserved quantities, which are extracted from the system and bath. In our case, following Jaynes \cite{jaynes1,Jaynes1957}, we take the `thermal bath' to be simply a collection of particles, each described by a generalised thermal state 
\begin{equation} \label{genthermal} 
\tau = e^{-(\beta_1 A_1 + \ldots + \beta_k A_k)}/\mathcal{Z}
\end{equation}
where $A_i$ are various conserved quantities,  $\beta_i$ are the associated inverse temperatures, and $\mathcal{Z}$ is the generalised partition function. Two things are important to note: First, the quantities $A_i$ may or may not commute, and even when they commute they may or may not be functionally dependent on one another. Second, and most importantly, energy need not be one of the conserved quantities or indeed play any role. Since energy is the generator of time evolution, such a thermal bath may not arise naturally by thermal equilibration, but have to be created externally (for example if the Hamiltonian is zero, then no evolution occurs). Yet as we will see, the thermodynamic flavour of the theory remains. 

The batteries are systems that can each store one of the conserved quantities $A_i$. In our paper we will consider the batteries either explicitly or implicitly, as explained later. The system can be an individual quantum particle. Finally, the actions that we allow to be performed must conserve either exactly or on average each of the quantities $A_i$, which is the content of the first law.  

A central result of standard thermodynamics -- the content of the second law -- is that if we have access only to a single thermal bath, it is impossible to extract energy, in the ordered form of work, out of it, that is $W := \Delta E^{\batt} \leq 0$, where $\Delta E^{\batt}$ is the change in the average energy of the battery. We show that, in our case, there is no limit on how much of any single conserved quantity $A_i$ can be extracted, even though we have access only to a single generalised bath. More precisely, there is no limit on $W_{A_i} := \Delta A_i^{\batt}$. There is however a global limit. 

In particular, to each conserved quantity we can associate an entropic quantity $\beta_i A_i$ (the entropic nature of this quantity will be explained later). We will show that these  quantities can be almost perfectly interconverted for one another inside the bath. As a result, due to the first law (conservation of $A_i$ between bath and battery) the only constraint on the $W_{A_i}$ given just a thermal bath is that 
\begin{equation}  \label{justbath}
\sum_i \beta_i W_{A_i} \leq 0.
\end{equation}

In standard thermodynamics the second law also says that if we have access to a system out of equilibrium with respect to the bath, then we can extract work, but we are limited by the change in free energy of the system, $W \leq -\Delta F^{\sys}$. In our case, we define an entropic quantity, the \emph{free entropy} of the system relative to the generalised bath, $\tilde{F}^{\sys}$, 
\begin{equation}
\tilde{F}_\s := \sum_i \beta_i \langle A_i^\s\rangle - S_\s
\end{equation}
and show that
\begin{equation} \label{bathplussystem} 
\sum_i \beta_i W_{A_i} \leq -\Delta \tilde{F}_\s.
\end{equation}
We will show, with a minimal number of assumptions, that we can implement any trade-offs between conserved quantities satisfying Eq. \eqref{justbath} using the bath, and extract any combination of $W_{A_i}$ satisfying  Eq. \eqref{bathplussystem} from a system, up to an arbitrarily small deficit due to the finite nature of the protocols. In particular, if all the conserved quantities commute we will give explicit protocols that works for both implicit or explicit batteries, assuming exact conservation of the $A_i$. For more general non-commuting quantities, we will obtain the same results for implicit batteries, or explicit batteries with average conservation, but leave open the question of how to deal with strict conservation of non-commuting quantities when considering explicit battery systems.

\section{The generalised thermal state} In this section, we consider in more detail the generalised thermal state given in Eq.~\eqref{genthermal} \cite{jaynes1,Jaynes1957}. 

We begin by recalling that there are two ways to define the thermal state -- by maximizing the von Neumann entropy given appropriate constraints, or by minimising the free energy. We start with the former.
Consider a system in state $\sigma$ with Hamiltonian $H$ and average energy $\langle H\rangle := \tr[H\sigma] = \overline{E}$. There are many states $\sigma$ that have this particular average energy; the thermal state is the state which maximizes the entropy  $S(\sigma)=-\tr(\sigma \ln \sigma)$, subject to the average energy constraint. Solving the maximization problem we get
$
 \tau(\beta) = \frac{1}{\mathcal{Z}}e^{-\beta(\overline{E}) H}
$
where $\mathcal{Z}$ is the partition function and the inverse temperature $\beta$ is implicitly determined by the average $\overline{E}$.

In our framework we need the generalisation of this idea to the case of multiple conserved quantities. In particular, we consider $k$ quantities $A_i\,,i \in \{1, \cdots, k\}$ and place no restrictions on the relations between them: they may or may not commute; when they commute they may or may not be functionally dependent on one another\footnote{An example of two commuting and functionally dependent quantities are the Hamiltonian $H$ and angular momentum $L$, where $H = L^2/2I$. In this case the average of one does not uniquely determine the average of the other, however the range of admissible values is constrained, i.e. such that $|\langle L \rangle| \leq \sqrt{2I\langle H \rangle }$. An example of two non-commuting conserved quantities is $L_x$ and $L_y$.}. The generalised thermal state $\tau(\beta_1,\ldots,\beta_k)$ is then the state which maximises the entropy $S$ subject to the constraint that the conserved quantities $A_i$ have average value $\la A_i\ra = \overline{A_i}$. It is found to be

\begin{defin}
\textbf{Generalised thermal state}\\
 \begin{align}\label{thermal}
\tau(\beta_1,\ldots,\beta_k) = \frac{ e^{- \sum_i \beta_i A_i}}{\mathcal{Z}}.
 \end{align}
 where, $\beta_i$ is the inverse temperature conjugate to $A_i$, and the generalised partition function is $\mathcal{Z} = \tr (e^{- \sum_i \beta_i A_i})$.
\end{defin}
Note that in general, each $\beta_i$ is a function of all of the averages $\overline{A_i}$. In the case that the $A_i$ commute, the proof is a simple generalisation of the standard proof. For non-commuting observables, the proof is  more involved \cite{Liu2007}.

The second way to define the thermal state (when only energy is conserved) is to fix the inverse temperature $\beta = \frac{1}{T}$ and ask for the density matrix that minimises the free energy $F(\rho) = \la H\ra - TS(\rho)$. The state which solves this optimisation has exactly the Gibbs form $\tau(\beta) = e^{-\beta H}/\mathcal{Z}$. Since $\beta$ is given, the average energy is now implicitly defined, in contrast to the case above, where the average energy was given and the inverse temperature derived.

The idea is to do the same in the case of multiple conserved quantities, and recover the generalised thermal state via a generalised free energy. However, in the standard definition of free energy the temperature is the constant multiplying the entropy, but since we have no notion of multiple entropies, we are not afforded a way of coupling all the inverse temperatures. This is easily overcome if instead of the free energy, we define $\tilde{F}(\rho) = \beta \la H\ra - S(\rho)$, and it is trivial to generalise this quantity to the case of multiple conserved observables.
\begin{defin}
\textbf{Free Entropy}\\
The free entropy of a system $\rho$ is
\begin{align}\label{free}
\tilde{F}(\rho) = \sum_i \beta_i \la A_i \ra- S(\rho),
\end{align}
\end{defin}

The free entropy is always defined w.r.t. a set of inverse temperatures $\beta_i$. The generalised thermal state is then the state that minimises $\tilde{F}$ with fixed $\beta_i$. For a complete proof of this fact, see the appendix. 

\section{Conceptual viewpoint}

As noted in the introduction, the effects presented in this paper suggest it may be worthwhile to revisit the basic concepts of thermodynamics. A key aspect of this is the conceptual shift from the free energy to the free entropy.

First, we would like to emphasise that the change from the usual free energy to the free entropy is not a simple mathematical manipulation, but marks a fundamental conceptual difference. Indeed, in the standard approach to considering multiple conserved quantities, such as when considering the grand canonical ensemble, one introduces the chemical potential $\mu$ such that the free energy becomes
\begin{equation}
F(\rho) = \langle H \rangle + \mu \langle N \rangle - T S (\rho)
\end{equation}
where $N$ is the particle number operator. In this way, energy is singled out as the privileged quantity, with the chemical potential acting as the `exchange rate' between particle number and energy (and in the same way temperature acts as the exchange rate between entropy and energy). We argue that there is no reason to single out the energy, or any other quantity for that matter. In fact, it is possible to conceive of situations in which everything is degenerate in energy, and thus where energy plays absolutely no role. We are thus lead to introduce the free entropy, which naturally and uniquely treats all quantities on an equal footing. 

A second argument for considering free entropy over the free energy is that the latter might give one incorrect intuition. Indeed, in the standard treatment, the free energy puts bounds on how much of the conserved quantity (energy) can be extracted, and one may be tempted to think that even when we have multiple conserved quantities thermodynamics is about the bounds which constrain the extraction of individual quantities. However, as we will show, this is not the case, and there are no such bounds. The only limitation is on the trade-off between the conserved quantities, and this is precisely governed by the free entropy. It is only in the standard case of a single conserved quantity that one can choose to consider the free energy, or the free entropy, with both constraining the amount of work that can be extracted.

We also note that the thermal state is the state which minimises the free energy only when the temperature is positive; if the temperature is negative the thermal state (at negative temperature) is instead the state which maximises the free energy. On the other hand, for all temperatures (positive or negative), the thermal state always minimises the free entropy. 

Finally,  we note that $\tilde{F}(\rho_\s) - \tilde{F}(\tau_\s)$ is equal to the relative entropy between $\rho_\s$ and $\tau_\s$, which highlights the entropic nature of $\tilde{F}$. 

Following on from the introduction of the free entropy, one can go a step further. Since the quantities $\beta_i \langle A_i \rangle$ all appear alongside the entropy in the definition of the free entropy, it suggests that they may be thought of as \textit{entropic quantities}. Note that this is true even in the standard case, where energy is the only conserved quantity; there we may think of $\beta \langle H \rangle$ as an entropic quantity. 

Importantly, the sum of these entropic quantities of the batteries is the object onto which the second law of thermodynamics applies. It says that the increase in this sum is constrained by the decrease in the free entropy of the system relative to the bath \eqref{bathplussystem}. 

\section{The setup}

The setup is similar in spirit to that of  previous works \cite{paul,Malabarba2015}. 
We consider the interaction of generalised thermal baths with quantum systems and batteries. There are a number of conserved quantities, $A_1$ to $A_k$, which may or may not commute or functionally depend on each other. The generalised thermal bath consists of an (unbounded) collection of systems, each of which is in a generalised thermal state as defined by Eq.~\eqref{thermal}. Any given protocol will involve only a finite set of systems in the bath, whose combined thermal state can be written as $\tau_\b(\beta_1,\ldots,\beta_k)$.
We  also want to consider an additional quantum system $\rho_\s$ that is both initially uncorrelated from and out of equilibrium with respect to the generalised bath, i.e. $\rho_\sb = \rho_\s \otimes \tau_\b(\beta_1,\ldots,\beta_k)$ and $\rho_\s \neq \tau_\s(\beta_1,\ldots,\beta_k)$. The main question we ask is how much of each of the conserved quantities can be extracted from the system (in conjunction with the bath, and stored in an associated battery).

In the interest of being clear, we proceed by concentrating on a scenario with only two conserved quantities, $A$ and $B$, since this already captures the majority of the physics contained in the general case of $k$ conserved quantities. 

In order to talk about the extraction of the conserved quantities there are two ways in which one can proceed: by either including battery systems implicitly, or explicitly, in the formalism. In the former case, one allows the global amount of each quantity stored in the system and bath to change, and define the changes as the amount of `$A$-type work' and `$B$-type work' that have been extracted from (or done on) the global system. The idea is that due to global conservation laws, when the $A_i$ of the system and bath changes, this change is compensated by a corresponding change to the external environment (the implicit battery). 

In the latter case, one introduces explicit battery systems which by definition only accept a single type of work (i.e. an $A$-type battery, and a $B$-type battery). Here, by definition the amount of $A$ stored in the $A$-type battery is the $A$-type work, and similarly for $B$. We enforce that the global amount of $A$ and $B$ stored in the system, bath and battery is constant, either strictly (the entire distribution is conserved) or on average. 

In the main text we consider the case of implicit batteries. We do this since dealing with implicit batteries simplifies the considerations and allows us to focus on what is arguably the most important part of the protocols, namely the interaction between system and bath. 
Obviously it is preferable to have the full protocol including batteries explicitly. In doing so there are many subtleties, which also arise in the case of standard thermodynamics. In particular we need to impose `no cheating' conditions that make sure that we do not make illegitimate use of batteries as sources of free entropy \cite{paul,johan2,Malabarba2015}. The danger stems from the fact that the batteries are systems out of equilibrium with respect to the bath. In the appendix we show how to include explicit batteries for a number of cases, as specified in Table \ref{table1}. 

\begin{table}
\begin{tabular}{cc|c|c}
                                                                                                                 &          & Commuting & Non-commuting \\ \hline
\multicolumn{1}{l|}{\multirow{2}{*}{\begin{tabular}[c]{@{}l@{}}Implicit\\ batteries\end{tabular}}}                  & 2nd law  & $\checkmark $         & $\checkmark $             \\ \cline{2-4} 
\multicolumn{1}{l|}{}                                                                                               & protocol & $\checkmark $         & $\checkmark $             \\ \hline
\multicolumn{1}{l|}{\multirow{2}{*}{\begin{tabular}[c]{@{}l@{}}Explicit batteries\\ (strict cons.)\end{tabular}}} & 2nd law  & $\checkmark $         & $\checkmark $             \\ \cline{2-4} 
\multicolumn{1}{l|}{}                                                                                               & protocol & $\checkmark $         & ?             \\ \hline
\multicolumn{1}{l|}{\multirow{2}{*}{\begin{tabular}[c]{@{}l@{}}Explicit batteries \\ (ave. cons.)\end{tabular}}}   & 2nd law  & $\checkmark $         & $\checkmark $            \\ \cline{2-4} 
\multicolumn{1}{l|}{}                                                                                               & protocol & $\checkmark $         & $\checkmark $*            
\end{tabular}
\caption{\label{table1} Summary of the results contained in the paper. The second law \eqref{eq:second law I} holds in all instances. * designates that the result holds only for explicit batteries with continuous spectra.  }
\end{table}

More concretely, when considering implicit battery systems the class of allowed transformations consists of all global unitary transformations $U$ on the system and bath. After such a transformation the global state is $\rho'_\sb = U(\rho_\s \otimes \tau_\b(\beta_A,\beta_B))U^\dagger$ with the reduced state of the system and bath given by the reductions, $\rho_\s' = \tr_\b [\rho'_\sb]$ and $\rho_\b' = \tr_\s[\rho'_\sb]$ respectively. We define the $A$-type and $B$-type work to be
\begin{equation}\label{eq:first laws}
\begin{split}
\Delta W_{A} &= -\Delta A_\s - \Delta A_\b \\
\Delta W_{B} &= -\Delta B_\s - \Delta B_\b \\
\end{split}
\end{equation}
where $\Delta A_\s = \tr[(A_\s (\rho'_\s - \rho_\s)]$, $\Delta A_\b = \tr[A_\b(\rho'_\b - \tau_\b)]$, and analogously for $\Delta B_\s$ and $\Delta B_\b$. Also, note that if our protocol involves multiple bath systems then $A_\b = \sum_i A_\b^{(i)}$, where $A_\b^{(i)}$ acts non-trivially only on bath system $i$, i.e. $A_\b$ is the sum of the local $A$ for each system (and analogously for $B_\b$). In \eqref{eq:first laws} we are equating the average change of $A$ and $B$, due to the unitary transformation, with the amount of $A$-type and $B$-type work that has been extracted from the system and bath. As such, our framework automatically incorporates the \emph{first law of thermodynamics} for each of the conserved quantities.

Finally, an additional unrelated problem, but which often plays an important role, concerns the precise structure of the bath. In usual treatments, we may consider particles in the bath which have any energy level spacing, such that their occupation probabilities can match any probabilities in the external system. This is used to construct efficient protocols. When considering other quantities than energy, we may be faced with quantities whose spectrum is fixed, such as angular momentum. Also, extra constraints or relationships may exist between the different conserved quantities. This results in additional difficulties. To address these, and to remain as general as possible, we will consider baths with a minimal amount of accessible structure.

\section{The second law}
Of great interest to us is the particular form that the second law of thermodynamics takes in the present setting. In the classical thermodynamic setting, the second law states that if one only has access to a thermal bath, then no work can be extracted, and that the maximal amount of work that can be extracted from a non-equilibrium system interacting with a thermal bath is bounded by the change in its free energy.

In our framework of multiple conserved quantities, we will see that the second law constrains the different combinations of conserved quantities which can be extracted from the system. In particular, we will show below that in the present framework, the amount of $A$-type work and $B$-type work that can be extracted is constrained such that
\begin{equation}\label{eq:second law I}
 \beta_A \Delta W_A + \beta_B \Delta W_B \leq -\Delta \tilde{F}_\s,
\end{equation}
where $\Delta \tilde{F}_\s = \tilde{F}(\rho'_\s) - \tilde{F}(\rho_\s)$. In the case where there is no system, or when the system is left in the same state, $\rho_\s' = \rho_\s$, then $\Delta \tilde{F}(\rho_\s) = 0$, and we obtain as a corollary
\begin{equation}\label{eq:second law II}
 \beta_A \Delta W_A + \beta_B \Delta W_B \leq 0.
\end{equation}
Equations \eqref{eq:second law I} and \eqref{eq:second law II} constitute the second law when one has multiple conserved quantities (with and without a system). 

To prove the second law \eqref{eq:second law I}, alongside the first laws \eqref{eq:first laws}, we will need to use two further formulas. First, since we restrict to unitary transformations the total entropy of the global system remains unchanged, $S(\rho'_\sb) = S(\rho_\sb)$, and from the fact that the system and bath are initially uncorrelated, along with sub-additivity, we have
\begin{equation}\label{eq:entropy}
\Delta S_\s + \Delta S_\b \geq \Delta S_\sb = 0
\end{equation}
where $\Delta S_\s = S(\rho'_\s) - S(\rho_\s)$, and analogously for $\Delta S_\b$ and $\Delta S_\sb$. Second, since the bath starts in the thermal state $\tau(\beta_A,\beta_B)$, which is a minimum of the free entropy (by definition), its free entropy cannot decrease during the protocol, thus
\begin{equation}\label{eq:free energy increase}
\Delta \tilde{F}_\b = \beta_A \Delta A_\b + \beta_B \Delta B_\b - \Delta S_\b \geq 0
\end{equation}
Now, eliminating all quantities on the bath, by substituting from the first laws \eqref{eq:first laws} and from  \eqref{eq:entropy}, we finally arrive at 
\begin{equation}
-\beta_A (\Delta A_\s + \Delta W_A) - \beta_B (\Delta B_\s + \Delta W_B) + \Delta S_\s \geq 0
\end{equation}
which, after re-arranging and identifying terms is straightforwardly seen to be \eqref{eq:second law I}, as desired. Thus, the first law, in conjunction with the lack of initial correlations (and sub-additivity), and the extremality of the generalised thermal state imply in a direct manner that systems obey a second law of the form given. We note that the proof does not rely on any particular properties of $A$ and $B$, which need not even commute. 

At this point it is worth briefly returning to the issue of implicit versus explicit batteries. If explicit batteries are included, then the unitary operations have to be extended to act on the system, bath and explicit batteries. Crucially, \eqref{eq:entropy}, and as a consequence the second law \eqref{eq:second law I}, can be shown to hold when we are careful to avoid cheating via batteries. Details are provided in the appendix.

In the remaining we will study to what extent we can saturate \eqref{eq:second law I} and \eqref{eq:second law II}, depending upon the properties of  the conserved quantities (whether they commute or not), whether we consider implicit or explicit batteries, and whether we consider strict or average conservation.

\section{Commuting observables}\label{s:trading}
We will now specialise to the case of commuting observables, and show how the second law can be saturated, both in terms of trading resources, and when extracting resources from a non-equilibrium system. 

In order to remain as general as possible, we want to assume as little as possible about the structure of the generalised thermal bath. What we will require is that there exists a system in the bath (of which we can take arbitrarily many copies) with $d\geq 3$ states, $\ket{a_i, b_i}$, for $i \in  \{0,1,2, \ldots d-1\}$, which are the joint eigenstates of $A_\b$ and $B_\b$ such that $A_\b\ket{a_i,b_i} = a_i\ket{a_i,b_i}$ and $B_\b\ket{a_i,b_i} = b_i\ket{a_i,b_i}$. We then need two requirements. First, that the observables should be sufficiently different. In particular, that 
\begin{align}\label{e:requirement}
\frac{a_1-a_0}{b_1-b_0} \neq \frac{a_2-a_0}{b_2-b_0}
\end{align}
which amounts to saying that $A$ and $B$ should not be affinely related to each other, in which case they should not be thought of as different quantities. Second, that in the thermal state the joint eigenstates should not have the same populations. In particular, it must be that
\begin{equation}\label{eq:no conspiracy}
\begin{split}
x &:= \beta_A(a_1-a_0) + \beta_B(b_1-b_0) \neq 0, \\
y &:= \beta_A(a_2-a_0) + \beta_B(b_2-b_0) \neq 0.
\end{split}
\end{equation}
If both $x$ and $y$ simultaneously vanish, then all three states have the same populations, in which case the system looks maximally mixed in this subspace. When trading quantities inside the bath, this will be the only problematic case. However, when we come to processing non-equilibrium systems, we will require simultaneously $x \neq 0$ and $y \neq 0$ in order for the bath to have enough structure to allow us to approach reversibility. We will also see that, depending on how close to reversible we want to be, we will have to exclude a small set (non-dense and of measure zero)  of joint values for $x$ and $y$ which are rationally related, as will be explained later.

\subsection{Trading resources}
We now consider the situation where we only have access to a generalised bath (and no external system), with the goal to show that we can perform a unitary transformation such that: (i)  its free entropy $\Delta \tilde{F}_\b = \beta_A \Delta A_\b + \beta_B \Delta B_\b$ (since $\Delta S_\b = 0$ by definition) changes by an arbitrarily small amount. (ii) the  changes $\Delta A_\b$ or $\Delta B_\b$ can be made arbitrarily large. In this case, we will say that we can exchange $A$ for $B$ in an essentially reversible manner. 

Therefore, let us consider that we have a $n$ copies of $\tau(\beta_A,\beta_B) = e^{-\beta_A A - \beta_B B}/\mathcal{Z} = \sum_i q_i \ket{a_i,b_i}\bra{a_i,b_i}$. It will be convenient to label states by the number of systems found in a given eigenstate, which we shall refer to as the occupation: We denote by $\ket{\mathbf{n},\alpha} \equiv \ket{n_0,n_1,\ldots,n_{d-1},\alpha} = P_\alpha \ket{a_0,b_0}^{\otimes n_0}\ket{a_1,b_1}^{\otimes n_1}\cdots\ket{a_{d-1},b_{d-1}}^{\otimes n_{d-1}}$, where $P_\alpha$ is a permutation operator, permuting the bath systems, labelled by $\alpha$, and $n_0 + n_1 + \ldots + n_{d-1} = n$. Now, we will consider only two states from the $d^n$ which are available, corresponding to
\begin{equation}\label{e:pair}
\begin{split}
\ket{\mathbf{n},\alpha} &=\ket{n_0,n_1,n_2,n_3,\ldots, n_{d-1},\alpha}  \\
\ket{\mathbf{n}',\alpha'}&=\ket{n_0',n_1',n_2',n_3,\ldots, n_{d-1},\alpha'}
\end{split}
\end{equation}
i.e. such that it is only the occupations of the first three levels differ between these states. As such, we have the constraint that $n_0 + n_1 + n_2 = n_0' + n_1' + n_2'$. The key step in our protocol is to apply a swap operation between these two states, whilst leaving all other states unchanged. A direct calculation shows that the change in the average value of each quantity of interest is
\begin{align}
\Delta A_\b &= \Delta q\left(a_{10}\Delta n_1 + a_{20}\Delta n_2\right)  \\
\Delta B_\b &= \Delta q\left(b_{10}\Delta n_1 + b_{20}\Delta n_2\right) \\
\Delta F_\b &= \Delta q\big(x\Delta n_1 + y\Delta n_2\big)  \label{Fbeqn}
\end{align}
where $\Delta n_k = (n'_k - n_k)$, $a_{k0} = (a_k - a_0)$, $b_{k0} = (b_k - b_0)$, $x$ and $y$ are as defined in \eqref{eq:no conspiracy}, and 
\begin{align}
\Delta q &= \left(1-\left(\tfrac{q_1}{q_0}\right)^{\Delta n_1}\left(\tfrac{q_2}{q_0}\right)^{\Delta n_2}\right)\prod_{i=0}^{d-1}q_i^{n_i} 
\end{align}
is the difference in populations between the two states. Given $y \neq 0$, we can rewrite Eq. \eqref{Fbeqn} as
\begin{equation} \label{Fbeqn2}
\Delta \tilde{F}_\b = y\, \Delta q\,\Delta n_1 \left( \frac{x}{y}  + \frac{\Delta n_2}{\Delta n_1} \right)
\end{equation} 
Now, for arbitrary $\Delta n_1$, we can find an integer $m$ such that 
$
m / \Delta n_1< x/y \leq (m+1)/ \Delta n_1 
$ 
 Setting $\Delta n_2 = -m$ in Eq. \eqref{Fbeqn2}, we obtain 
 \begin{equation} 
0 < \Delta \tilde{F}_\b \leq  y \Delta q.
\end{equation} 
Hence $\Delta \tilde{F}_\b$ can be made as small as desired by making $\Delta q$ arbitrarily small (which can be achieved by increasing $n_0$)\footnote{Note that it is important that $\Delta \tilde{F}_\b \neq 0$, as the thermal state is the unique  state which minimises $\tilde{F}_\b$, so this would require that the bath is left completely unchanged, giving $\Delta A_\b=\Delta B_\b=0$}. On the other hand, we find that the relative change in the conserved quantities $\Delta A_\b/\Delta \tilde{F}_\b$ and $\Delta B_\b/\Delta \tilde{F}_\b$ are
\begin{equation}
\begin{split}
\frac{\Delta A_\b}{\Delta \tilde{F}_\b} &= \frac{a_{20}}{y}\left(1 + \frac{\beta_B b_{20}}{y}\left(\frac{a_{10}}{a_{20}}-\frac{b_{10}}{b_{20}}\right)\left(\frac{x}{y}+\frac{\Delta n_2}{\Delta n_1}\right)^{-1}\right) \\
\frac{\Delta B_\b}{\Delta \tilde{F}_\b} &= \frac{b_{20}}{y}\left(1 + \frac{\beta_A a_{20}}{y}\left(\frac{b_{10}}{b_{20}}-\frac{a_{10}}{a_{20}}\right)\left(\frac{x}{y}+\frac{\Delta n_2}{\Delta n_1}\right)^{-1}\right) \\
\end{split}
\end{equation}
In both cases, the final term satisfies $ (x/y + \Delta n_2/\Delta n_1)^{-1} \geq \Delta n_1$ and can hence be made as large as desired by increasing $\Delta n_1$. This means that the magnitude of $\Delta A_\b/\Delta \tilde{F}_\b$ and $\Delta B_\b/\Delta \tilde{F}_\b$ will become arbitrarily large. The sign of  $\Delta A_\b$ will depend on the other constants, but can be modified if desired by  choosing $m$ such that  $(m-1) / \Delta n_1 \leq x/y < m/ \Delta n_1$ above. Note also that if $y =0$ but $x\neq 0$ we can construct an equivalent proof with the roles of $x$ and $y$ swapped. 

Finally, by repeating the above protocol a sufficient number of times, one can trade arbitrary amounts of the conserved quantities from a generalised bath by sacrificing an arbitrarily small amount of free entropy. In particular, to achieve $\Delta A_\b^{\total} \geq \eta$ with $\Delta \tilde{F}^{\total}_\b \leq \epsilon$, one can perform the protocol above ($\epsilon/\Delta \tilde{F}_\b$) times, with $\Delta A_\b/\Delta \tilde{F}_\b \geq \eta / \epsilon$.

At this stage a few comments are necessary. First, it is important to stress that the protocol relies on a minimal amount of structure in the observables $A$ and $B$ and the bath: it requires that there exist many copies of a bath system with a $3$-dimensional subspace where the action of the operators are not trivially related (by a shift and rescaling), and that the state is not maximally mixed in this subspace. Moreover,  each bath system is taken to be identical, with no additional parameters necessary (i.e. we do not require a family of different $A_\b$'s and $B_\b$'s, like the families of Hamiltonians  considered in \cite{paul}).
Second, this protocol is a only a proof-of-principle demonstration that trade-offs can be enacted. In particular, we paid no attention to the number of generalized thermal states necessary. If one cared about minimising the resources utilised, then clearly the above protocol would not be used, and more efficient ones would be sought. Finally, we note that the above analysis generalises beyond two conserved quantities to the general case of $k$ quantities. In this case, 
condition \eqref{e:requirement} much hold pair-wise for all quantities.

\subsection{Extracting resources from a single quantum system}\label{s:extracting}
In the previous subsection we showed that  arbitrarily good interconversions can be enacted  given access only to a generalised bath, we now move on to the scenario of having a quantum system out of equilibrium with respect to the bath. Our goal is to show that we can saturate the second law given by Eq.~\eqref{eq:second law I}  arbitrarily well, -- i.e. that we can extract  conserved quantities from a non-equilibrium system such that $\beta_A \Delta W_A + \beta_B \Delta W_B$ is as close as desired to the system's  decrease  in free entropy.

Let us consider that we have a state $\rho_\s$, which in terms of its eigenstates and eigenvalues is given by $\rho_\s = \sum_i p_i\ket{\psi_i}\bra{\psi_i}$, and by convention we take the eigenvalues to be ordered, $p_{n+1}\le p_n$. In general the eigenbasis of the state will not coincide with the joint eigenbasis of the conserved quantities $A$ and $B$. The first step is to pre-process the system, to bring it to a diagonal form in this basis. To do so we will not interact with the  bath, but simply apply the unitary 
\begin{equation}\label{eq: U diagonal}
U_\s  = \sum_i \ket{a_i,b_i}\bra{\psi_i},
\end{equation}
on the system such that $\sigma_\s = U_\s\rho_\s U_\s^\dagger = \sum_i p_i \ket{a_i,b_i}\bra{a_i,b_i}$. Due to the first laws \eqref{eq:first laws}, we have that $\Delta W_A = -\Delta A_\s$ and $\Delta W_B = -\Delta B_\s$. Moreover, since the entropy of the system did not change, $\Delta \tilde{F}_\s = \beta_A \Delta A_\s + \beta_B \Delta B_\s$, and we have immediately
\begin{equation}
-\Delta \tilde{F}_\s = \beta_A \Delta W_A + \beta_B \Delta W_B
\end{equation}
i.e. in this state, unsurprisingly, we have a change in the batteries which coincides with the free entropy change of the system, and saturates \eqref{eq:second law I}. Note that in the case of explicit battery systems, the transformation \eqref{eq: U diagonal} cannot be perfectly performed at the level of the system, since $[U_\s,A_\s] \neq 0$ and $[U_\s,B_\s] \neq 0$. Nevertheless, one can implement a joint unitary on the system and batteries such that at the level of the system the transformation $U_\s$ can be approximated arbitrarily well, as long as the battery systems are in appropriate states, similarly  to the case of standard quantum thermodynamics  \cite{johan2,Malabarba2015}. Full details can be found in the appendix. 

Now, having bought the system to diagonal form, we want to consider a transformation $\sigma_\s \rightarrow\sigma'_\s  =\sum_i p_i^\prime \ket{a_i^s,b_i^s}\bra{a_i^s,b_i^s}$ in which only two levels of the system change their populations by a small amount\footnote{We denote the eigenvalues of $A_\s$ and $B_\s$ by $a_i^\s$ and $b_i^\s$, to differentiate them from the eigenvalues of $A_\b$ and $B_\b$, which we will similarly denote by $a_i^b$ and $b_i^b$.}. Namely, we would like to perform $p_1^\prime = p_1 +\delta p,\; p_0^\prime = p_0 - \delta p$. We can implement such changes by finding two levels in the bath whose ratio of populations are close to $p_0/p_1$ and then swapping the two-dimensional subspaces of the system and bath. To carry out any such transformation requires a finely spaced set of different population ratios in the bath.  

Let us now consider whether the simple bath systems we have considered so far can provide such possibilities. As before, we bring in a collection of $n$ generalised thermal states, $\tau(\beta_A,\beta_B)$ and consider only the two states \eqref{e:pair}, $\ket{\mathbf{n},\alpha}$ and $\ket{\mathbf{n}',\alpha'}$. We shall denote the populations of  these states by $q_{\mathbf{n}}$ and $q_{\mathbf{n}'}$ respectively. The global unitary transformation we will apply is the swap operator between the two-dimensional subspaces of the system and bath, and the identity everywhere else. That is, the operation that performs
\begin{equation}
\ket{a_0,b_0}\ket{\mathbf{n}',\alpha'} \leftrightarrow \ket{a_1,b_1}\ket{\mathbf{n},\alpha}
\end{equation}
whilst leaving all other states unchanged. By performing this transformation the population that is shifted between the states of the system, which coincides with the population that is shifted between the states of the bath, is $\delta p = (p_0 q_{\mathbf{n}'} - p_1 q_{\mathbf{n}})$. Now, the changes in the conserved quantities of the system and bath are found to be
\begin{align}
\Delta A_\s &= \delta p (a_1^s - a_0^s),& \Delta B_\s &= \delta p (b_1^s - b_0^s) \nonumber \\
\Delta A_\b &= -\delta p (a_{\mathbf{n}'}^b - a_{\mathbf{n}}^b),& \Delta B_\b &= -\delta p (b_{\mathbf{n}'}^b - b_{\mathbf{n}}^b).
\end{align}
The change in the entropy of the system and the bath are also found to be\footnote{Note that we use $\frac{p_0'}{p_1'}$ rather than $\frac{p_0}{p_1}$ here, because we can always take $p_1' \neq 0$.}
\begin{align}
\Delta S_\s &= \delta p \log \frac{p_0'}{p_1'} + O(\delta p^2) \nonumber \\
\Delta S_\b &= -\delta p \log \frac{q_{\mathbf{n}}}{q_{\mathbf{n}'}} + O(\delta p^2)
\end{align}
We want to achieve $\Delta S_\s + \Delta S_\b = O(\delta p^2)$, which requires that $\log (q_\mathbf{n} / q_{\mathbf{n}'} ) = \log (p_0' / p_1') + O(\delta p)$. To see when this condition is satisfied, we first use the explicit form of the probabilities, which show that $q_\mathbf{n}/q_{\mathbf{n}'} = \exp(-x\Delta n_1 - y\Delta n_2)$. Hence we require $\log(p_1'/p_0') - x\Delta n_1 - y \Delta n_2 = O(\delta p)$. Since $p_0'$ and $p_1'$ are arbitrary, our requirement is that $x \Delta n_1 + y \Delta n_2$ should be able to come within $O(\delta p)$ of any number. As we can rescale $ \Delta n_1$ and $ \Delta n_2$ by an arbitrary integer, it is sufficient to obtain $0 < x \Delta n_1 + y \Delta n_2 \leq O(\delta p)$. If $x$ and $y$ are not rationally related this is always possible. However if $x/y$ is rational, and given in reduced form by $u/v$ (where $u$ and $v$ are co-prime integers), then we need 
\begin{equation} 
0 < \frac{y}{v} (u  \Delta n_1 + v \Delta n_2) \leq O(\delta p).
\end{equation} 
From number theory, one can always find $ \Delta n_1$ and $ \Delta n_2$ such that the term in brackets is 1 (or -1), hence we need $|y / v| \leq O(\delta p)$. For a fixed desired accuracy $\epsilon$ (of order $O(\delta p)$), and fixed $y$, this rules out a finite number of $x$ values for which $x/y$ is a rational with a small denominator. Extending this to the $x-y$ plane, we find that we can achieve the desired accuracy everywhere except for a non-dense set of measure zero. 

Returning to the entropies, with the above in place, $\Delta S_\s + \Delta S_\b = O(\delta p^2)$, i.e. the system and bath remain essentially uncorrelated after the transformation. Finally, we use once again the fact that the generalised thermal state is a minimum of the free entropy. This implies that the changes in population, of order $O(\delta p)$, change the free entropy only to second order, $\Delta \tilde{F}_\b = O(\delta p^2)$. Putting everything together we have
\begin{align}
\Delta \tilde{F}_\b &= \beta_A \Delta A_\b + \beta_B \Delta B_\b - \Delta S_\b, \nonumber \\
 &= -\beta_A (\Delta W_A + \Delta A_\s) - \beta_B (\Delta W_B + \Delta B_\s) \nonumber \\
&\quad\quad\quad+ \Delta S_\s + O(\delta p^2), \nonumber \\
&= -\Delta \tilde{F}_\s - (\beta_A \Delta W_A + \beta_B \Delta W_B) + O(\delta p^2).
\end{align}
Thus, since the left-hand-side is $O(\delta p^2)$, it must be that 
\begin{equation}
\beta_A \Delta W_A + \beta_B \Delta W_B = -\Delta \tilde{F}_\s + O(\delta p^2)
\end{equation}
i.e. that up to a correction of order $O(\delta p^2)$, the combination of conserved quantities extracted, which themselves are order $O(\delta p)$, matches the change in free entropy of the system. Thus, by composing $O(1/\delta p)$ of such transformations, we can implement a protocol which transforms $\rho_\s \to \tau_\s(\beta_A,\beta_B)$, whereby in each stage the population changes between two states by order $O(\delta p)$, and such that 
\begin{equation}
\beta_A \Delta W_A + \beta_B \Delta W_B = -\Delta \tilde{F}_\s + O(\delta p).
\end{equation}
Therefore, by taking $\delta p$ sufficiently small we can approach the reversible regime, whereby the change in free entropy of the system matches the dimensionless combination of conserved quantities extracted. Combining this protocol with the protocol from the previous section, involving only the generalized bath and the batteries, we can obtain any combination of extracted conserved quantities.

At this stage, a few comments are again in order. First, it is important to stress that our assumptions only changed by a small amount relative to the previous section. In particular, we need a minimal amount of extra structure  in the bath, such that it is useful to process arbitrary individual systems out of equilibrium. Second, as previously, the protocol presented generalises in a straightforward manner to the case of $k$ mutually commuting conserved quantities $A_1$,\ldots $A_k$. 

Finally, note that the same protocol can also be used to perform efficient transformations between any two system states (where the final state is full rank), and not just to the thermal state. Our protocol also immediately gives an asymptotic protocol for the interconversion of states with no average work cost: the rate at which one can transform $\rho^{\otimes n}$ into $\sigma^{\otimes nR}$ is given by
\begin{equation}
R = \frac{\tilde{F}(\rho)-\tilde{F}(\tau(\beta_1,\ldots,\beta_k)}{\tilde{F}(\sigma)-\tilde{F}(\tau(\beta_1,\ldots,\beta_k)}
\end{equation} 
Here, one can simple run the protocol `forward' individually on $n$ copies of $\rho$, in order to obtain in the batteries $n(\beta_1 \Delta {W}_{A_1} + \ldots \beta_k \Delta {W}_{A_k})$, and create $\tau(\beta_1,\ldots,\beta_k)^{\otimes n}$. Then, on $nR$ copies run the protocol `backwards' to create $\sigma^{\otimes nR}$, having returned each battery so that finally it contains the same amount its associated quantity that it initially contained (on average). 

In the appendix we show how these results extend to the case of explicit batteries with either strict or average conservation. We also show how the protocol can be made robust to experimental imperfections -- i.e. without assuming precise knowledge of $\beta_A$ or $\beta_B$.

\section{Non-commuting observables}
In this section we will show that when considering implicit batteries, the results obtained in the previous section can easily be modified to also work for non-commuting observables. This also extends to explicit batteries with average conservation laws when the batteries have continuous spectrum. However, the same protocols do not obviously generalise to the case of explicit batteries with strict conservation. 

Whereas previously, by virtue of the commutativity of the observables we could find a joint eigenbasis $\ket{a_i,b_i}$ that was used in our explicit protocols, that is no longer the case for non-commuting observables. Nevertheless, the generalised thermal state is diagonal in the basis of $\beta_A A + \beta_B B$,
\begin{equation}
e^{-\beta_A - \beta_B B}/\mathcal{Z} = \sum_i q_i \ket{i}\bra{i}
\end{equation}
Although to each eigenstate we can no longer associate an eigenvalue for $A$ or $B$, we can still associate an \emph{average value}, 
\begin{align}
\langle a \rangle_i &:= \bra{i}A\ket{i}, & \langle  b \rangle_i &:= \bra{i}B\ket{i}.
\end{align}
The main point is that all of our previous results hold if instead of joint eigenstates $\ket{a_i,b_i}$, with eigenvalues $a_i$ and $b_i$, we use the eigenstate $\ket{i}$ with average values $\langle a \rangle_i$ and $\langle b \rangle_i$ throughout. 

The only subtlety that arises is the structure we need from the bath. We still only need to use three distinct eigenstates, $\ket{0}$, $\ket{1}$ and $\ket{2}$, but now the necessary structure relates to the average value of the conserved quantities in the eigenstates, First, it must be that 
\begin{equation}
\frac{\langle a \rangle_1 - \langle a \rangle_0}{\langle b \rangle_1 - \langle b \rangle_0} \neq \frac{\langle a \rangle_2 - \langle a \rangle_0}{\langle b \rangle_2 - \langle b \rangle_0}
\end{equation}
otherwise, at the level of the average values, the observables appear affinely related and therefore cannot be sufficiently distinguished to allow for trade-offs. Furthermore, we still need to be able to find eigenstates in the bath which differ in population, just as before. We can define the analogous quantities $\langle x \rangle$ and $\langle y \rangle$, and if they do not simultaneously vanish then the bath will not be maximally mixed in the subspace. Finally, in order to extract resources from systems out of equilibrium with respect to the generalised bath there must be sufficient structure such that any ratio of populations can be approximated well enough. Again, in complete analogy to the above, if $(\langle x \rangle / \langle y \rangle)$ is irrational, then we have sufficient structure. If on the other hand $(\langle x \rangle / \langle y \rangle)$ is rational, we will again have to exclude a small set of values of $\langle x \rangle$ and $\langle y \rangle$  (non-dense, of zero measure), for which our results will not hold.

\section{Conclusions}
In this work we have studied a generalisation of thermodynamics where there are multiple conserved quantities, where energy may not even be part of the story. We have been interested in what form the second law takes, and showed that it is no longer about restrictions on individual extractable quantities, but rather about the allowed ways that the conserved quantities can be traded-off for one another. Indeed, we found that we can extract as much of any individual conserved quantity as desired, as long as the other conserved quantities are appropriately consumed in the process, with the second law dictating how much of the others are necessarily consumed. In particular, we were lead to introduce a dimensionless generalisation of the free energy, which we termed the free entropy, that is the central quantity which appears in the second law and dictates the allowed trade-offs. Moreover, given access to any quantum system out of equilibrium with respect to the generalised bath, we showed that its free entropy change bounds the combination of conserved quantities that can be extracted. 

Our results hold both for commuting and non-commuting observables, and with the desire to remain as general as possible we made only very mild assumptions about the bath. Indeed, we assumed as very little about the relationship between the conserved quantities or their individual structure. The one case which remains open for future research is the case of non-commuting observables, with explicit batteries and strict conservation of the conserved quantities. Although the protocols presented for saturating the second law do not appear to generalise to this case, we do not know whether entirely different constructions will be able to achieve this goal. 

\section*{Acknowledgements}
We thank J. Hannay for many discussions while carrying out this research. We also thank S. Massar for useful comments. YG, PS and SP acknowledge support from the European Research Council (ERC AdG NLST). SP Acknowledges support from a Royal Society Wolfson Merit Award. AJS acknowledges support from the Royal Society and FQXi via the SVCF. RS acknowledges support from the Swiss National Science Foundation (Starting grant DIAQ). YG, AJS, RS and PS acknowledge support from the COST Action MP1209 “Thermodynamics in the quantum regime”.

\section*{Note Added}
While completing this work we learnt that N. Yunger Halpern, P. Faist,
J. Oppenheim and A. Winter were independently working on related issues \cite{YungerHalpern2015}. While in the final stages of writing, the related work of M. Lostaglio, D. Jennings and T. Rudolph appeared on the arXiv \cite{Lostaglio2015c}. 

\bibliographystyle{apsrev4-1}
\bibliography{ConservedRefs}
\appendix
\begin{center}
\textbf{Appendices}
\end{center}
\section{The generalised thermal state}
In this section we prove that the generalised thermal state, in our framework of multiple conserved observables, can be obtained by either minimising the free entropy or by maximising the von Neumann entropy.

Minimising the free entropy is relatively simple, regardless of the relationship between the observables. We use the definition of free entropy from the main paper (Eq. 3) with $n$ conserved observables labelled by $A_i, \; i\in \{1, \cdots, n\}$,
\begin{align}
\tilde{F}(\rho) &= \sum_i \beta_i \langle{A_i}\rangle- S(\rho).
\end{align}
where $\beta_i$ are the inverse temperatures corresponding to each observable, $\langle A_i\rangle = \tr[A_i \rho]$ are the averages of each conserved quantity, and $S(\rho) = -\tr[\rho \log \rho]$ is the von Neumann entropy. We relabel the linear combination of observables as a single operator $R = \sum_i \beta_i {A}_i$ so that
\begin{align}
\tilde{F}(\rho)&= \tr [{R} \rho]  - S(\rho).
\end{align}

We first note that the state which minimizes this function must be diagonal in the eigenbasis of $R$ (If it was not, one could simply de-cohere the state in this basis, resulting in a state with an identical value for $\tr [{R} \rho]$  but higher entropy). If the occupation probability of the eigenstate with eigenvalue $R_i$ is denoted by $p_i$, one can  perform Lagrangian optimization to extremise 
\begin{equation} 
\mathcal{L}(\rho) = \sum_i R_i p_i + \sum_i p_i \log p_i + \lambda (\sum_i p_i -1) 
\end{equation} 
where $\lambda$ is a Lagrange multiplier to obtain 
\begin{align}
\rho = \frac{e^{-R}}{\mathcal{Z}} = \frac{e^{-\sum_i\beta_i A_i}}{\mathcal{Z}}
\end{align}
where the partition function is $\mathcal{Z} = \tr[e^{-\sum_i\beta_i A_i}$. Our aim is now to show that we arrive at the same form of solution if we maximise the von Neumann entropy subject to the averages of the conserved observables being fixed. There are two cases to consider here, depending on whether the observables $A_i$ commute with each other or not.

\emph{Commuting observables. ---} In this case, we wish to maximise the entropy subject to the constraints that all $n$ observables commute with one another i.e $[A_i, A_j]=0, \quad\forall  i, j\in \{1, \cdots, n\}$ and that each observable has some fixed average value on the system $\overline{A}_i$.

Since all of the observables commute, there exists a common eigenbasis, and the state that maximizes the entropy will necessarily be diagonal in this basis. (Otherwise, as above, one could de-cohere the state in the common eigenbasis, resulting in a state with the identical averages $\overline{A}_i$ but higher entropy).

The problem can be expressed as a Lagrangian optimization subject to the constraints that the state is normalised and the average observable quantities are constant
\begin{align}
\mathcal{S}(\rho) &= -\tr[\rho \log \rho]  - \sum_i \beta_i(\tr[\rho A_i]-\overline{A}_i) + \lambda (\sum_i p_i -1)\nonumber\\
&=-\sum_j p_j \log p_j -  \sum_{i} \beta_{i}(\sum_j p_j m_j^i - \overline{A}_i)\nonumber \\
& \quad  +  \lambda (\sum_j p_j -1)
\end{align}
where we interpret $\lambda$ and $\beta_i$ as the Lagrange multipliers and $m_j^i$ is the $j-$th eigenvalue of the $i$-th observable. By solving $\frac{\partial S}{\partial p_i} =0$ we find the solution is 
\begin{align}
\rho = \frac{e^{-\sum_i \beta_i A_i }}{\mathcal{Z}},
\end{align}

\emph{Non-commuting observables. ---} For the case of non-commuting observables we will have to be more careful as we can no longer diagonalise them in the same basis. \\
First we imagine that the average of the conserved observables are given to us, i.e. we have the numbers $\overline{A}_i $ for $ i \in \{1, \cdots, n\}$. Now we consider a state of the form
\begin{align}\label{thermalstate}
\rho = \frac{e^{-\sum_i \beta_i A_i}}{\mathcal{Z}}.
\end{align}
Since we know the operators $A_i$ we can compute their averages on the state $\rho$ as functions of the inverse temperatures $\beta_i$
\begin{align}\label{step1}
A_i  (\rho)= \tr [\rho A_i]  = f_i (\boldsymbol\beta) \qquad \forall i \
\end{align}
where $\boldsymbol\beta $ is the vector of inverse  temperatures $(\beta_1, \cdots, \beta_n)$. For the average quantities that we have been given, we now solve for the $\beta_i$
\begin{align}\label{step2}
  f_i (\boldsymbol\beta)= \overline{A}_i   \qquad\qquad\quad \forall i
\end{align}
The fact there there always exist solutions is non-trivial. In particular it implies that for any given set of average values $\{\overline{A}\}_i$ (here we implicitly assume compatible average values), there exist corresponding temperatures $\beta_i$ such that a state of the form (\ref{thermalstate}) pertains to these averages. This result was first proved by Jaynes \cite{jaynes1} and subsequently via a different method by Kai \cite{kai}. Solving these equations, we find the particular solutions $\beta_i^*$. We now define a new function $\tilde{F}^*$ which acts on density operators
\begin{align}
\tilde{F}^* (\mu) = \sum_i \beta_i^*\langle {A_i}\rangle_\mu  - S(\mu)
\end{align}
Next,  consider a density operator $\gamma$ with the properties that it has average values $\langle{A_i}\rangle_\gamma$ equal to $\overline{A}_i$ and that it also maximises the entropy $S(\gamma)$. Then $\tilde{F}^*(\gamma)$ is 
\begin{align}
\tilde{F}^* (\gamma) &= \sum_i \beta_i^*\langle{A_i}\rangle_\gamma - S(\gamma)\\
&= \sum_i \beta_i^* \overline{A}_i - S(\gamma)
\end{align}
We now consider the unique density operator $\sigma$ that minimises $\tilde{F}^*$. It can be obtained via the method in the first section and is simply 
\begin{align}
 \sigma &= \frac{e^{-\sum_i \beta_i^* A_i }}{\mathcal{Z}}\label{consider}
\end{align}
However, from eqs.(\ref{step1}) and (\ref{step2}) we observe that the state with inverse temperatures $\beta_i^*$ and  the form given in eq.(\ref{consider}) has averages ${\overline{A}_i}$ (since the non-trivial solutions to a linear system are unique), which implies
\begin{align}
\tilde{F}^*(\sigma) = \sum_i \beta_i^* \;{\overline{A}_i}  - S(\sigma).
\end{align}
Since $\tilde{F}^*(\rho)$ is at a minimum, it implies that $S(\sigma)$ is maximum, thus
\begin{align}
\tilde{F}^*(\sigma) = \tilde{F}^*(\gamma) \quad\Longrightarrow \quad \sigma = \gamma 
\end{align}
and we see that the state which maximises the entropy and has averages $\{\overline{A}_i\}_i$ is indeed the generalised thermal state. 

\section{Explicit batteries: allowed operations and the second law}
In the main text we presented the proof of our claim for the case that the work storage systems (the batteries) were implicit. Here we present an extended framework in which the batteries are explicit.

\indent Our setup is much the same as in the main text: we have a bath $b$, system $s$ and in addition, two batteries, which we call weights $w_A$ and $w_B$. We model the battery systems as weights in the most general sense,  whose value for each observable is given by a position observable (for energy, the height of the weight corresponds to the stored energy, but note that these weights need not be gravitational). If the observable has a discrete spacing such as angular momentum, then the weight may be a ladder with discrete spacing, but otherwise we take it to be continuous.\\\indent

 The value of observable $A$ on the weight is proportional to the position operator $A_{w_A} =c_a\hat{x}_a$, where $c_a$ is a constant of appropriate units in order to recover the correct dimensions for the quantity $A$, and we define the work by $\Delta W_A = \Delta A_{w_A}$. At this stage it is also useful to define the translation operator
 \begin{align}\label{translate}
 \Gamma_{w_A}^\epsilon= \exp(-i \epsilon \hat{p}_a )
\end{align} 
  where $\epsilon$ reflects the amount of translation of the weight and $\hat{p}_a$ corresponds to the operator canonically conjugate to the position operator $\hat{a}$ (i.e. the momentum). The translation operator effects the following transformation on unnormalised position states of the pointer
$
\Gamma_{w_A}^\epsilon \ket{x_a}= \ket{x_a+\epsilon} 
$. Analogously we make the same definitions for quantity $B$. 
Differences in the average position of the weight before and after the protocol allows us to read off the change in the work of that quantity.\\\indent
Our intention here is to remain as general as possible, whilst eliminating the possibility of `cheating' by bringing in resources from outside this framework (such as external sources of work or free energy), or making use of objects within the framework for a purpose other than intended (for example, by using the batteries as a cold reservoirs in generalised heat engines). We make four assumptions on our scheme:
\begin{enumerate}[label=\Roman*]
\item \label{rule1} We assume that the \textit{battery systems are independent} of one another and only accept and store one type of conserved quantity. As such, each quantity is assigned its own battery system.
\item\label{rule2}
The set of allowed operations will consist of \textit{global unitaries} on the bath, system and weights, $U$, which conserve $A$ and $B$. Using rule \ref{rule1} we have
\begin{equation}
\begin{split}
[U, A_\b + A_\s + A_{w_A}] &=0\label{e:strict}\\
 [U, B_\b + B_\s + B_{w_B}]& = 0
\end{split}
\end{equation}

In this way, we impose the first laws of thermodynamics for any initial state:
\begin{equation}
\begin{split}
\Delta A_\b + \Delta A_\s + \Delta A_{w_A} &= 0\\
\Delta B_\b + \Delta B_\s + \Delta B_{w_B} &= 0 
\end{split}
\end{equation}
We choose to study unitaries as opposed to more general completely positive (CP) maps in order not to use  external ancillas in non-thermal states as sources of energy or angular momentum.

\item \label{rule3} We assume \textit{translational invariance} of the weights to reflect the fact that only displacements in the position on the ladders are important. This implies that all unitaries $U$ should commute with translation operators on each weight.
\item \label{rule4} Finally, we assume that all four bodies  are \textit{initially uncorrelated} and start in the product state $\rho_\s \otimes \tau_\b \otimes \rho_{w_A}\otimes \rho_{w_B}$. 
\end{enumerate}

The proof of the second law in the presence of explicit batteries follows the same logic as the implicit proof up to a few subtleties. 
\begin{theo}\label{Uentropy}
All unitary evolutions $U$ which are weight-translation invariant cannot decrease the entropy of the system and bath.
\begin{align}
[U, \Gamma_{w_A}^{\epsilon_1} ] = [U, \Gamma_{w_B}^{\epsilon_2} ]= 0\quad \Longrightarrow \quad \Delta S(\rho_\sb)\ge 0
\end{align}
where the translation operators $\Gamma$ are defined in eq.(\ref{translate}).
\end{theo}

\textit{Proof. } Using the definition in eq.(\ref{translate}) we associate two momentum-like variables, conjugate to the positions of the pointers, for quantities $A$ and $B$. For clarity we let $p_a = p$ and $p_b = \phi$. 
We argue that since the weights are both translationally invariant, this means that $[U,p]=[U,\phi] = 0$. Any unitary with this property can be written
\begin{align}
U = \int dp \,d \phi \,V(p, \phi)\otimes \ket{p}\bra{p}\otimes \ket{\phi}\bra{\phi}
\end{align}

where the first element of the tensor product  $V(p, \phi) $ corresponds to a unitary operation on the combined system and bath (as a function of the variables conjugate to the positions $x_a$ and $x_b$ of the two weights) and the second corresponds to a projector onto the un-normalised momentum eigenstates $\ket{p}$ and $\ket{\phi}$ of weights $w_A$ and $w_B$.
By rule \ref{rule4} the weights are initially uncorrelated from the bath and system we can write the initial state (in density matrix form) as $\rho_\sb\otimes\rho_{w_A}\otimes \rho_{w_B}$. 
We are interested in the post-measurement state of $\rho_\sb^\prime$. After applying the unitary and tracing out the battery systems, the state is

\onecolumngrid
\begin{align}
\rho_\sb^\prime &= tr_{w_Aw_B} \bigg(U(\rho_\sb \otimes \rho_w\otimes \rho_t) U^\dagger\bigg)\\
& = tr_{w_Aw_B} \bigg(\int dp \;dp^\prime\;d\phi \;d{\phi}^\prime \,  V({p, \phi} ) \rho_\sb V^\dagger ({p}^\prime, \phi^\prime) \otimes \bigg(  \ket{p}\bra{p} \rho_{w_A} \ket{p^\prime}\bra{p^\prime}\otimes \ket{\phi}\bra{\phi} \rho_{w_B} \ket{\phi^\prime}\bra{\phi^\prime} \bigg)  \bigg)\\
&=  \int dp \;d \phi \;  V({p,\phi} ) \rho_\sb V^\dagger ({p, \phi})\, \otimes\underbrace{\bigg( \bra{p}\rho_{w_A}\ket{p} \otimes\bra{\phi}\rho_{w_B}\ket{\phi} }_{\text{mixing terms}} \bigg)\\
&= \int dp \;d\phi \;  V({p,\phi} ) \rho_\sb V^\dagger ({p,\phi})\, \alpha(p)\nu(\phi)
\end{align}
\twocolumngrid
\noindent
where $\alpha(p)$ and $\nu(\phi)$ are the probability distributions for the momenta on the initial state of the weights. The system and bath therefore evolve via a mixture of unitary transformations. Due to the concavity of the entropy, and the fact that it is preserved under the unitary transformation $V(p, \phi)$, such evolutions can only increase the entropy of the system and bath
\begin{align}
S(\rho_\sb)&\le S\bigg(\int dp\, d\phi\;  V(p, \phi)\rho_\sb V(p,\phi)^\dagger \alpha(p)\nu(\phi)\bigg) \nonumber\\
&= S(\rho_\sb^\prime)\\
\Longrightarrow 0 &\le \Delta S(\rho_\sb) \qquad\square\label{entincrease}
\end{align}

We now calculate the von Neumann entropy change of the bath and system. Following rule \ref{rule4} the bath and system are initially uncorrelated, thus their initial entropy is simply the sum of their individual entropies. The unitary we implement may be entangling, and therefore correlations may form between the bath and system during the protocol. Using the result in eq.(\ref{entincrease}) and the fact that entropy respects subadditivity, we have that
\begin{align}
\Delta S_\b + \Delta S_\s &\ge \Delta S_\sb \ge 0 \\
\Delta S_\b &\ge - \Delta S_\s\label{bathenropy}
\end{align}
Following the line of thought from the main text, we now argue that the free entropy of the bath can only  increase 
\begin{align}
\Delta \tilde{F}_\b = \beta_A \Delta A_\b  + \beta_B \Delta B_\b - \Delta S_\b \ge 0.
\end{align}
Using rule $\ref{rule3}$ and the entropy relation (\ref{bathenropy}) we arrive at the second law
\begin{align}
\beta_A (-\Delta A_\s - \Delta W_A)  + \beta_B (-\Delta B_\s - \Delta W_B) - \Delta S_\s &\ge 0 \\
\Longrightarrow\quad \beta_A \Delta \tilde{W}_A  +\beta_B \Delta \tilde{W}_B \le -\Delta \tilde{F}_\s &
\end{align}
Note that if rule $\ref{rule2}$ is changed to the case of only average quantity conservation (i.e. Eq.(\ref{e:strict})) is dropped) then proof of the second law still holds. Thus our result is universal for both strict and average quantity conservation. 

\section{Generalised work extraction details for the case of explicit weights}

In this section we show how the  protocols for trade-offs between conserved quantities and generalised work extraction in the main paper can be extended from implicit to explicit batteries, in the case of commuting observables. This follows closely the approach in \cite{Malabarba2015,johan2}.

Any protocol on the system and bath in the implicit battery framework can be represented by a total unitary transformation $U$ (which may be the product of several unitary steps). We can write this transformation as 
\begin{equation} 
U = \sum_{ij} U_{ij} \ket{i}\bra{j} 
\end{equation} 
where the basis states $\ket{i}$ are joint eigenstates of $ A_\s + A_\b$ and $B_\s + B_\b$ with eigenvalues $a_i$ and $b_i$ respectively. 

In the explicit battery framework described in the previous section, a general unitary $U$ would not be allowed as it does not strictly conserve the quantities $A$ and $B$. However, we can instead perform the unitary 
\begin{align}
\tilde{U}&= \sum_{ij} U_{ij} \ket{i}\bra{j} \otimes \Gamma_{w_A}^{a_j - a_i}\otimes \Gamma_{w_B}^{b_j - b_i}.
\end{align}
which commutes with $A_{total} = A_\s + A_\b + A_{w_A}$ and $B_{total} = B_\s + B_\b + B_{w_A}$. We will now show that for appropriate initial states of the weights (in particular very broad coherent states with momentum approximately zero) $\tilde{U}$ has approximately the same effect on the system and bath as $U$. Due to the first laws, the work extracted into the weights will then be approximately the same as in the case with implicit batteries. Furthermore, this approximation can be made as good as desired, and the protocol does not degrade the state of the weights for use in further protocols. 

In the momentum representation, this can be written
\begin{align} \label{Uform}
\tilde{U}= \int \int V(p, \phi) \otimes \ket{p}\bra{p}\otimes \ket{\phi}\bra{\phi} dp\,d\phi
\end{align}
where
$
V(p, \phi) = \sum_{ij} U_{ij} e^{-ip (a_j - a_i)}e^{-i\phi (b_j - b_i)} \ket{i}\bra{j}.
$

Ideally, in order to implement $U$, we want $V(p, \phi) = V(0, 0)=U$, which corresponds to a very narrow wavefunction for the momentum of $w_A$ and $w_B$. To show that this can always be done, we trivially extend the proof of Malabarba et al. in Theorem 1 of \cite{Malabarba2015} to the case of two battery systems and show that the state of the system $\rho^\prime_\s$ after the global unitary will remain close in trace distance to that of the desired local evolution $U \rho_\s U^\dagger$. Let $\rho^\prime_\sb =\tr_{w_A\,w_B} (\tilde{U} \rho_\sb \otimes \rho_{w_A}\otimes \rho_{w_B} \tilde{U}^\dagger) $, we thus want to show 
\begin{align}\label{e:closetrace}
\bigg|\bigg|  &\tr_{w_A\,w_B} (\tilde{U} \rho_\sb\otimes \rho_{w_A}\otimes \rho_{w_B} \tilde{U}^\dagger) - U\rho_\sb U^\dagger  \bigg|\bigg|\nonumber\\
& = \bigg|\bigg|\int\int dp \;d\phi\braket{p}{\rho_{w_A}}{p}\braket{\phi}{\rho_{w_B}}{\phi} V(p, \phi) \rho_\sb V^\dagger(p,\phi)\nonumber\\
& \qquad- U \rho_\sb U^\dagger\bigg|\bigg|\; \le \epsilon
\end{align}
for an arbitrary $\epsilon>0$. $\braket{p}{\rho_{w_A}}{p}$ and $\braket{\phi}{\rho_{w_B}}{\phi}$ are well defined probability distributions of the two weights, which we will denote $\mu_{w_A}(p)$ and $\mu_{w_B}(\phi)$. Since $V(p, \phi)$ is a continuous function of its variables, then there always exists $\delta, \Delta$ such that
\begin{align}\label{e:max}
\max_{(p,\phi) \in I} \bigg|\bigg|V(p, \phi) \rho_\sb V^\dagger(p,\phi) -V(0,0) \rho_\sb V(0,0)^\dagger\bigg|\bigg| <\frac{\epsilon}{2}
\end{align}
where $I =\{(- \delta, \delta), (- \Delta, +\Delta)\}$. We now choose initial weight states such that the probability distributions $\mu_{w_A}(p)$ and  $ \mu_{w_B}(\phi)$ satisfy

\twocolumngrid

\begin{align}
\int_{-\delta}^{\delta}\mu_{w_A} (p) \, dp  \int_{-\Delta}^{\Delta}  \mu_{w_B} (\phi) d \phi \,   \geq 1 - \frac{\epsilon}{2}.\label{e:prob2}
\end{align}
Substituting eqs.(\ref{e:max}) -- (\ref{e:prob2}) into eq.(\ref{e:closetrace}) one arrives at the result, 
\begin{align}
\bigg|\bigg|  \tr_{w_A\,w_B} (\tilde{U} &\rho_\sb\otimes \rho_{w_A}\otimes \rho_{w_B} \tilde{U}^\dagger) - U \rho_\sb U^\dagger  \bigg|\bigg| \leq \epsilon 
\end{align}
as desired. Note that $\mu_{w_A} (p)$ and $\mu_{w_B} (\phi)$ are not changed by the protocol, due to the form of Eq. \eqref{Uform}, and hence the weights can be reused in future protocols without being degraded.  

For the work extraction protocol in particular, the total $\tilde{U}$ will be equal to a product of unitaries for each individual step, $\tilde{U} = \tilde{U}_1\tilde{U}_2$. As an explicit example, the unitary for the first step, in which $\rho_\s$ is rotated into the joint eigenbasis of $A$ and $B$ is given by 
\begin{align}
\tilde{U}_1&= \sum_{ij} c_{ij}^*\ket{a_i, b_i}\bra{a_j, b_j}_\s\otimes \mathbb{I}_\b \otimes \Gamma_{w_A}^{a_i - a_j}\otimes \Gamma_{w_B}^{b_i - b_j},
\end{align}
where  $\ket{\psi_i} = \sum_j c_{ij}\ket{a_j, b_j}$. The unitary which swaps a two-dimensional subspace of the bath and system is given by
\begin{align}\label{e:swap}\nonumber
\tilde{U}_2 = \mathbb{I}_\sb&\otimes  \mathbb{I}_{w_A}\otimes \mathbb{I}_{w_B} \qquad\qquad\qquad\qquad\qquad\qquad\qquad\\
+ &\bigg(\ket{\mathbf{n^\prime}{\alpha^\prime}, 0 }\bra{ \mathbf{n}\alpha,  1}_\sb\otimes \Gamma^{\epsilon_1}_{w_A}\otimes \Gamma^{\epsilon_2}_{w_B} \nonumber \\
& \quad +\ket{ \mathbf{n}\alpha, 1 }\bra{ \mathbf{n^\prime}{\alpha^\prime}, 0}_\sb\otimes \Gamma^{-\epsilon_1}_{w_A}\otimes \Gamma^{-\epsilon_2	}_{w_B}\bigg) \nonumber\\
-& \bigg((\ket{ \mathbf{n}\alpha, 1 }\bra{ \mathbf{n}\alpha, 1}_\sb + \ket{ \mathbf{n^\prime}{\alpha^\prime}, 0 }\bra{ \mathbf{n^\prime}{\alpha^\prime}, 0}_\sb)\otimes \mathbb{I}_{w_A}\otimes \mathbb{I}_{w_B}\bigg)\qquad\qquad\qquad
\end{align}
where  $\ket{\mathbf{n}\,\alpha} = \ket{n_0, n_1, n_2, n_3, \cdots  n_d, \alpha}$ and \\ $\ket{\mathbf{n^\prime}\alpha^\prime}=\ket{n_0^\prime, n_1^\prime, n_2^\prime, n_3, \cdots  n_d, \alpha^\prime}$ are the two states in occupation notation which we choose from the bath. In order to obey strict quantity conservation the weights must shift by the difference in the quantity gap in the system and bath, i.e.  $\epsilon_1 = ((a^s_1 -a^s_0) -(a^b_{\mathbf{n^\prime}} - a^b_{\mathbf{n}}))  $ and similarly $\epsilon_2 = ((b^s_1 -b^s_0) -(b^b_{\mathbf{n^\prime}} - b^b_{\mathbf{n}}))  $, where $a^s_i, b^s_i$ denote the eigenvalues of $A_\s, B_\s$ and  $a^b_i, b^b_i$ denote those of $A_\b, B_\b$.

\section{A robust protocol in the case of experimental uncertainty }
We present a protocol for work extraction from a system with multiple conserved observables in conjunction with a generalised thermal bath, which is robust even in the case that we have uncertainty in the temperatures of the baths. 
We wish to extract some amount of $
W_A$ and $W_B
$ and in order to do this we must implement the swap operation in eq.(\ref{e:swap}).As such we wish to match the ratio of probabilities $\tfrac{p^\prime}{p}$ in the system $\rho_\s$ with the ratio of probabilities $\tfrac{q_{\mathbf{n}}^\prime}{q_{\mathbf{n}}}$ in the bath $\tau(\beta_A, \beta_B)$. 

Specifically, the subscripts $\mathbf{n}$ and $\mathbf{n}^\prime$ in the bath refer to the particularly chosen occupation states (level distributions) $\mathbf{n}$ and $\mathbf{n}^\prime$ of the bath, which we will swap to implement the protocol. $\mathbf{n} = (n_0, n_1, n_2,n_3\cdots, n_d)$ and $\mathbf{n}^{\prime}=(n_0^{\prime},n_1^{\prime},n_2^{\prime}, n_3, \cdots, n_d)$ such that $\mathbf{n} - \mathbf{n}^\prime = ((n_0 - n_0^\prime), (n_1 - n_1^\prime), (n_2 - n_2^\prime), 0\cdots, 0)$.
For commuting observables $A, B$,  the bath probabilities take the following form:
\begin{align}\label{e:general}
q_i \propto e^{-(\beta_A a_i + \beta_B b_i)}
\end{align}
where $a_i (b_i)$ are the eigenvalues of the observable $A\,(B)$. Since we have taken the tensor product of $n$ thermal states $\tau(\beta_A, \beta_B)$, the ratio of probabilities between the two selected levels is simply
\begin{align}
\frac{q_{\mathbf{n}}}{q_{\mathbf{n}^\prime}} &= \frac{\prod_{i =0}^{d-1} q_i^{n}}{\prod_{i =1}^{d-1} q_i^{n^\prime_i}} \\
&= q_0^{n_0 - n_0^\prime }q_1^{n_1-n_1^\prime}q_2^{n_2-n_2^\prime}\\
&= \bigg( \frac{q_0}{q_1}\bigg)^{n_1^\prime - n_1} \bigg( \frac{q_0}{q_2}\bigg)^{n_2^\prime - n_2}
\end{align}
where we have used particle conservation $\sum_{i = 0}^2 n_i = \sum_{i=0}^2 n_i^\prime$ to eliminate the term $(n_0 - n_0^\prime)$ in the last line. Substituting for the general form of the probabilities from eq.(\ref{e:general}) we have 
\begin{align}\label{e:ratio}
\frac{q_{\mathbf{n}^\prime}}{q_{\mathbf{n}}} &= e^{-(x\Delta n_1 + y \Delta n_2)} \\
&=e^{-{y \Delta n_1}\bigg(\tfrac{x}{y} + \tfrac{\Delta n_2}{\Delta n_1}\bigg)}
\end{align}
where $x = (\beta_A (a_1 - a_0) + \beta_B(b_1 - b_0))$, $y = (\beta_A (a_2 - a_0) + \beta_B (b_2 - b_0))$  and $\Delta n_1 = (n_1^\prime - n_1), \;  \Delta n_2 = (n_2^\prime - n_2)$. 
We would like to show that for the quantity in eq.(\ref{e:ratio}) we can match any possible ratio given to us from the system $\tfrac{p^\prime}{p}$, which is equivalent to demanding
\begin{align}\label{e:small}
\bigg|{{y \Delta n_1}\bigg(\tfrac{x}{y} + \tfrac{\Delta n_2}{\Delta n_1}\bigg)} \bigg|<\epsilon
\end{align}
where $\epsilon$ is a constant of $O(\delta p)$. 

If we can achieve $\epsilon$ sufficiently small, then we can cover $\mathbb{R}^+$ well,  in the sense that we can come as close as desired to reproducing any number $0<\tfrac{p^0}{p^1}<\infty$ by scaling $\Delta n_1$ and $\Delta n_2$ by   $k$ (where $k \in \mathbb{Z}$). 
\begin{align}
0<\bigg((\tfrac{q_{\mathbf{n}^\prime}}{q_{\mathbf{n}}} )^k  = e^{-{{y k\Delta n_1}(\tfrac{x}{y} + \tfrac{\Delta n_2}{\Delta n_1})} }> e^{-k \epsilon} \bigg)< \infty
\end{align}
\begin{figure}[h!]
\begin{center}
\includegraphics[scale=0.30]{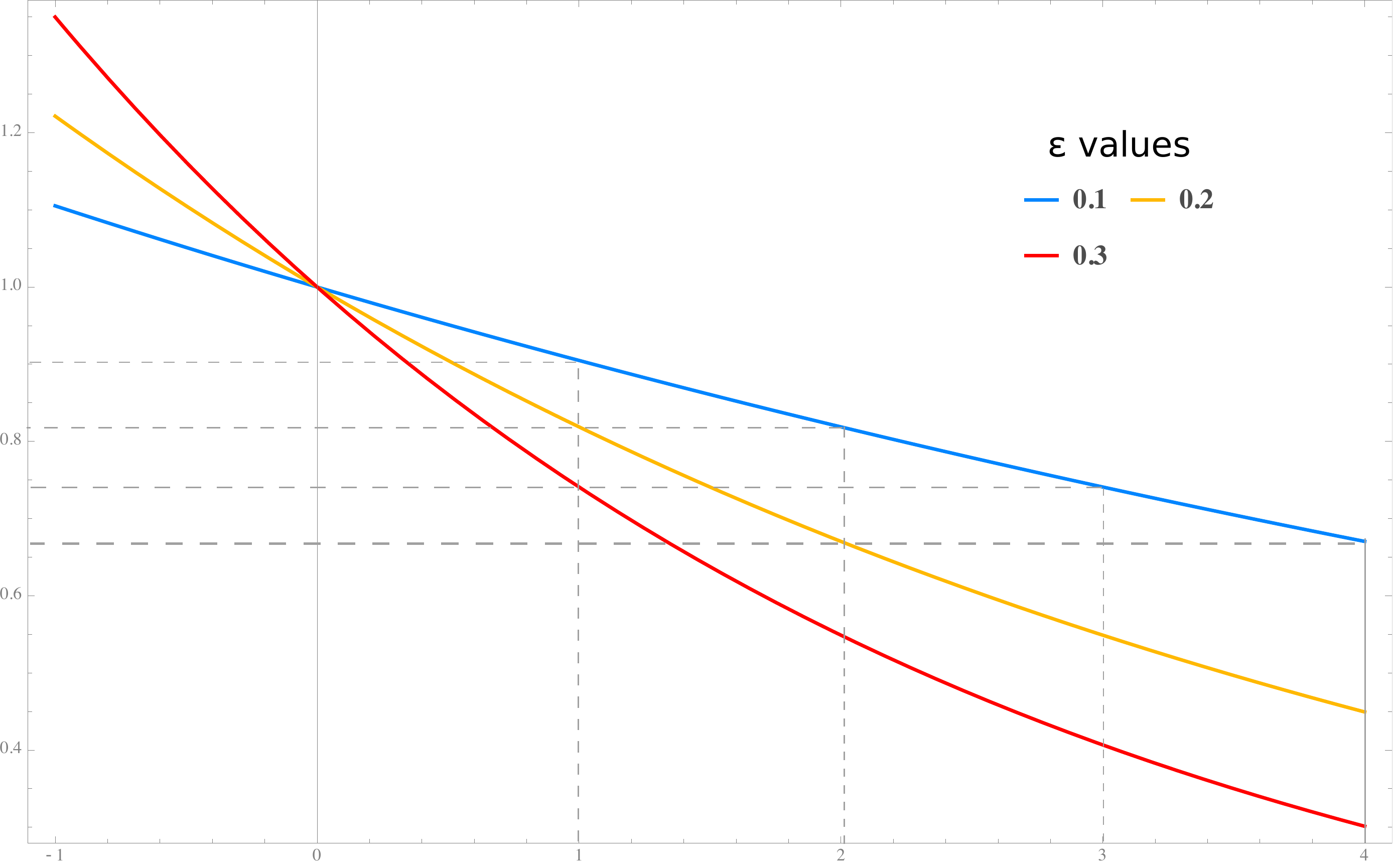}
\caption{A graph of $\tfrac{q_{\mathbf{n}^\prime}}{q_{\mathbf{n}}}= e^{- k\epsilon}$ for different values of $\epsilon$. As $\epsilon$ gets smaller, the values the function can attain for integer $k$ become closer together ($y$-axis intercepts), until they are sufficiently dense to reproduce any real number. }
\end{center}
\end{figure}
 We now present a robust method that will choose $\Delta n_1$ and $\Delta n_2$ such that eq.(\ref{e:small}) holds, even in the case of uncertainty due to the imprecision of the measuring apparatus. 

\indent  We begin by making the reasonable assumption that the experimenter measures values which are rational -- for instance because the measuring apparatus displays a finite string of decimal digits. The experimenter measures the temperatures of the thermal bath $\beta_A, \beta_B$ and computes $\tfrac{x}{y}$, specifying the uncertainty with $\delta$. They then find $\tfrac{x}{y} = \tfrac{u}{v}$ in its reduced form, such that $u, v$ are relative primes. For simplicity, we will consider the case where $0< x <y$, in which case $0 < \frac{x}{y} < 1$ (the other cases follow similarly).

The exponent we wish to make small becomes 
\begin{align}\label{e:regime1}
\bigg| y\Delta n_1\bigg(\tfrac{u}{v} + \tfrac{\Delta n_2}{\Delta n_1}\bigg)\bigg| = \bigg|\tfrac{y}{v} ({u \Delta n_1} + {v\Delta n_2}) \bigg|
\end{align}
The best we can do is to appeal to number theory: since $u$ an $v$ are relative primes, by B\'ezout's lemma \cite{bezout}, the smallest (in magnitude) non-zero value of $u\Delta n_1+v \Delta n_2$ is 1, and there exists a pair $\{\Delta n_1,\Delta n_2\}$ for which this is true. For this choice of pair, the quantity in eq.(\ref{e:regime1}) is automatically less than $\epsilon$ if $\frac{y}{v} <\epsilon$. What we find is that in the case that $\tfrac{y}{v}>\epsilon$, we do not know how to minimise this quantity because our best method fails. Thus, for fixed $\epsilon$, there is a finite set of points $\{\tfrac{p}{q}\}^F$ that are excluded from our protocol. This set is precisely the Farey sequence of order $\floor*{\frac{y}{\epsilon}}$, (where $\floor*{\bullet}$ denotes the floor function: the largest integer not greater than $\bullet$).

\begin{defin} [Farey Sequence] ``A Farey sequence of order $n$ is the ascending series of irreducible fractions between 0 and 1 whose denominators do not exceed $n$''. \cite{numtheory}. For example the 	 sequence of order 5 is $F_5= \{\frac{0}{1}, \frac{1}{4}, \frac{1}{3}, \frac{1}{2}, \frac{2}{3}, \frac{3}{4}, \frac{4}{5}, \frac{1}{1} \}$.
\end{defin}
\begin{figure}[h!]
\begin{center}
\includegraphics[width=\columnwidth]{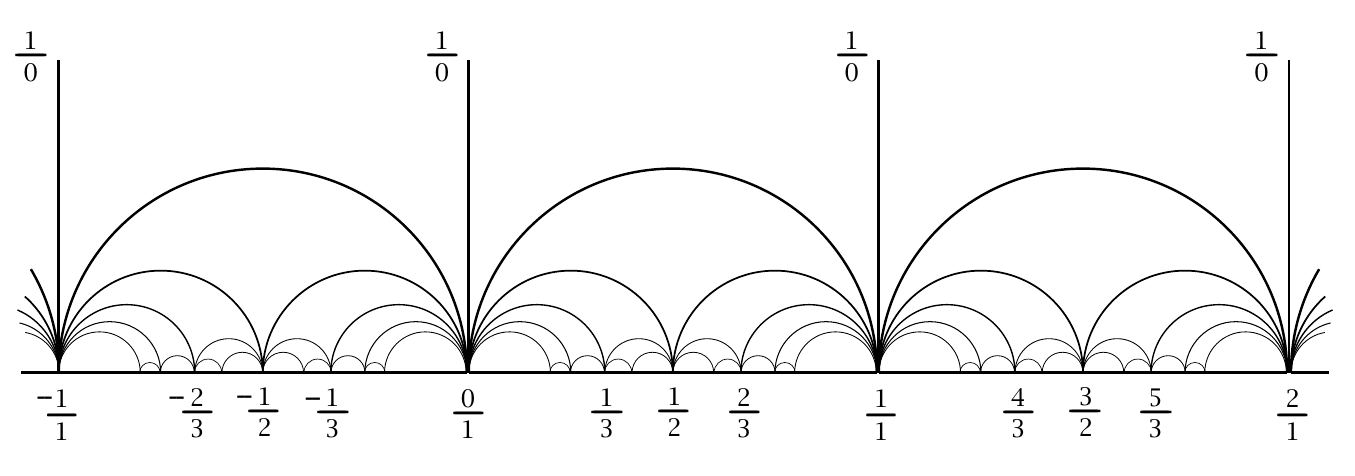}
\caption{The Farey sequence of order $5$. Image credit \cite{farey}.}
\end{center}
\end{figure}

We proceed by constructing a robust protocol around these excluded points. The experimenter measures $\tfrac{x}{y}$, computes $\tfrac{u}{v}$ and identifies the closest rational $\tfrac{u^*}{v^*}$ from the Farey sequence of order $\floor*{\tfrac{y}{\epsilon}}$ to the computed point.
It then suffices to choose $\{\Delta n_1=u^*,\Delta n_2=-v^*\}$. With this choice
\begin{align}
	\Mod{x\Delta n_1 +y\Delta n_2 } &= \Mod{yv^* \left( \tfrac{x}{y} - \tfrac{u^*}{v^*} \right)} \\
		&< \epsilon \;\;\quad \emph{iff} \;\;\; \Mod{\tfrac{x}{y} - \tfrac{u^*}{v^*}} < \tfrac{\epsilon}{yv^*}
\end{align}
For this choice, the quantity $\Mod{x\Delta n_1 +y\Delta n_2 } \le \epsilon$ for the interval $I_{u^*/v^*} = \left(\tfrac{u^*}{v^*} - \tfrac{\epsilon}{yv^*}, \tfrac{u^*}{v^*} + \tfrac{\epsilon}{yv^*} \right) - \{\tfrac{u^*}{v^*}\} $.

This method will work for the case that all the intervals overlap, otherwise there would be regions for which the protocol didn't work. The fact that the intervals overlap is the subject of the next theorem.
\begin{theo}[Farey intervals overlap]
The union of all intervals around each member of the Farey sequence of order $\floor*{\tfrac{y}{\epsilon}}$ cover the unit interval on which the Farey sequence is defined. 
\begin{align}
\bigcup_i I_{u/v_i} > [0, 1]
\end{align}
where ${u}/{v}_i$ is the $i$-th element of the Farey sequence and 
 $
I_{u/v} = \left(\frac{u}{v} - \frac{\epsilon}{yv}, \frac{u}{v} + \frac{\epsilon}{yv} \right) - \{\frac{u}{v}\}
$ is the interval. 
\end{theo}

\noindent\textit{Proof.} To prove that the collection of intervals around all bad rationals $\frac{u}{v}$ covers the real line, it suffices to prove that the neighbouring intervals in this collection intersect. Consider the rational number $\frac{u^\prime}{v^\prime}$ that is the next rational number (i.e. the neighbour) of $\frac{u}{v}$ in the Farey sequence of order $\floor*{\frac{y}{\epsilon}}$. Then the corresponding interval is $I_{u^\prime/v^\prime} = \left(\frac{u^\prime}{v^\prime} - \frac{\epsilon}{yv^\prime}, \frac{u^\prime}{v^\prime} + \frac{\epsilon}{yv^\prime} \right) - \{\frac{u^\prime}{v^\prime}\}$. Comparing the supremum of the interval about $\frac{u}{v}$ and the infimum of the interval about $\frac{u^\prime}{v^\prime}$, and using the properties of neighbours in a Farey sequence, one has that
\begin{align}
	\text{sup} \; \left( I_{u^\prime/v^\prime} \right) - \text{inf} \; \left( I_{u/v} \right) &= \tfrac{u^\prime}{v^\prime} - \tfrac{u}{v} - \tfrac{\epsilon}{y} \left( \tfrac{1}{v^\prime} + \tfrac{1}{v} \right) \\
		&= \tfrac{1}{vv^\prime} - \tfrac{\tfrac{\epsilon}{y}(v+v^\prime)}{vv^\prime}\label{Farey},
\end{align}
where we have used the property that if $\frac{u}{v}$ and $\frac{u^\prime}{v^\prime}$ are neighbours in a Farey sequence, then $\frac{u^\prime}{v^\prime} - \frac{u}{v} = \frac{1}{vv^\prime}$ \cite{numtheory}.

\begin{figure}
 \includegraphics[width = 0.48\textwidth]{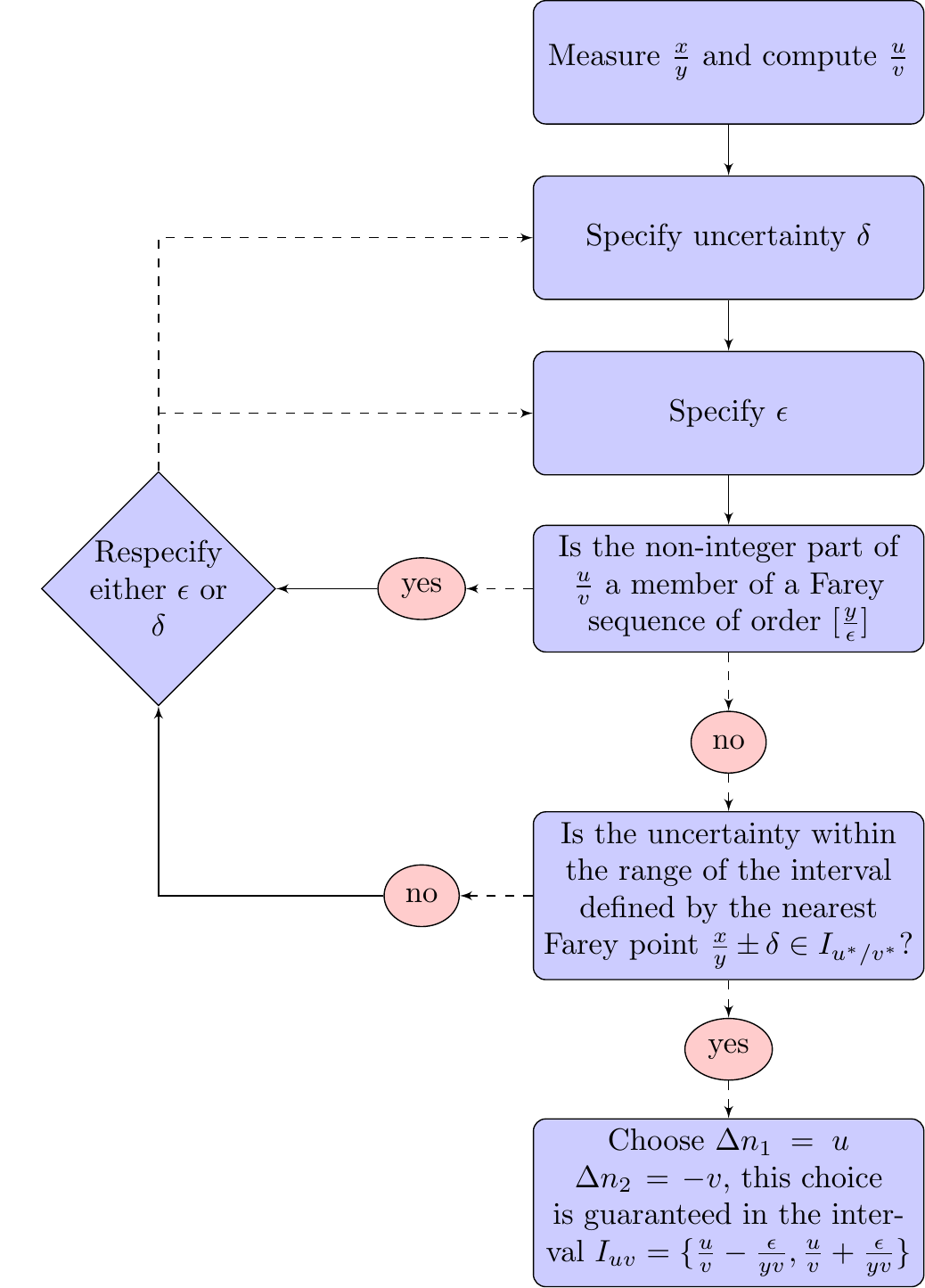}
 \caption{Flow chart showing a robust protocol for work extraction in the presence of measurement uncertainty.}
  \label{f:flow}
 \end{figure}

Furthermore, if $\frac{u}{v}$ and $\frac{u^\prime}{v^\prime}$ are neighbours in a Farey sequence of order $\floor*{\frac{y}{\epsilon}}$, then $v+v^\prime >\frac{ y}{\epsilon}$, else the mediant $\frac{(u+u^\prime)}{(v+v^\prime)}$ would also be in the Farey sequence of order $\floor*{\frac{y}{\epsilon}}$ which contradicts the assumption that $\frac{u}{v}$ and $\frac{u^\prime}{v^\prime}$ are neighbours. Therefore it follows from eq.(\ref{Farey}) that $\text{sup} \; \left(I_{u^\prime/v^\prime}\right) - \text{inf} \; \left(I_{u/v}\right) <0$. Thus the intervals overlap, and the union of the intervals around every rational number in the sequence covers the real line (less the excluded rational numbers) $\square$.\\\indent
Finally, the experimenter must verify that the uncertainty in their measurement, $\delta$, falls within the range of the Farey interval they have chosen, $\tfrac{x}{y}\pm\delta\in I_{u^*/v^*}$. If this criterion is not met, then the experimenter is obliged to respecify either $\epsilon$ or $\delta$, i.e. to respecify the fine-graining of the bath or make a more accurate measurement. \\
These arguments are easy to extend to the real line: the Farey sequence is translationally invariant on any unit interval and can be scaled by a constant as necessary;  the experimenter constructs the sequence of order $\floor*{\tfrac{y}{\epsilon}}$ between the integers that $\tfrac{x}{y}$ lies in, i.e. in the interval $ \floor*{\tfrac{x}{y}}<\tfrac{x}{y}<  \ceil*{\tfrac{x}{y}}$, (where $\ceil*{\bullet}$ denotes the ceiling function: the smallest integer not less than $\bullet$). 
\noindent The full protocol is summarised in Fig.(\ref{f:flow})

\end{document}